\documentclass[english,aps,article]{revtex4}
\usepackage[T1]{fontenc}
\usepackage[latin1]{inputenc}
\usepackage{babel}
\usepackage{graphics}
\usepackage{setspace}
\usepackage{amsmath}
\makeatletter

\providecommand{\LyX}{L\kern-.1667em\lower.25em\hbox{Y}\kern-.125emX\@}

\usepackage[T1]{fontenc}
\usepackage[latin1]{inputenc}
\usepackage{babel}
\usepackage{graphics}
\usepackage{graphicx}

\usepackage{epsfig}

\makeatletter

\usepackage[T1]{fontenc}
\usepackage[latin1]{inputenc}
\usepackage{babel}

\makeatletter

\usepackage[T1]{fontenc}
\usepackage[latin1]{inputenc}
\usepackage{babel}



\makeatother

\makeatother

\begin{document}

\title{Current relaxation in the Random Resistor cum Tunneling Network Model through First-Passage route : Regimes and Time-scales}

\author{Somnath Bhattacharya} 

       \email{somnath94347@gmail.com}

\affiliation{Department of Physics, Durgapur Govt College, J. N. Road, Paschim Bardhaman, $713201$, India.}

\vskip 1.0cm

\begin{abstract}

Numerically we study the bulk current relaxation in percolative Random Resistor cum Tunneling Network (RRTN) model through a first-passage route. The RRTN considers an extra semi-classical barrier-crossing process over a voltage threshold within a framework of classical RRN bond percolation model. We identify the different temporal regimes of relaxation and corresponding phenomenological time-scales, which fix up the extents of different regimes. These time-scales were previously identified in refs. \cite{relax-physicaA, aksubh}. We investigate on the distributions of these time-scales and observe that there exists a perfect correlation among them in the thermodynamic limit. We conclude that there exists a single time-scale which controls the RRTN dynamics. The variation of mean first-passage time .vs. system size seems to be due to sub-diffusive motion of charge carrier through the network. 

\end{abstract}  

\maketitle

\section{Introduction}

\begin{figure}
\resizebox*{7cm}{7cm}{\rotatebox{270}{\includegraphics{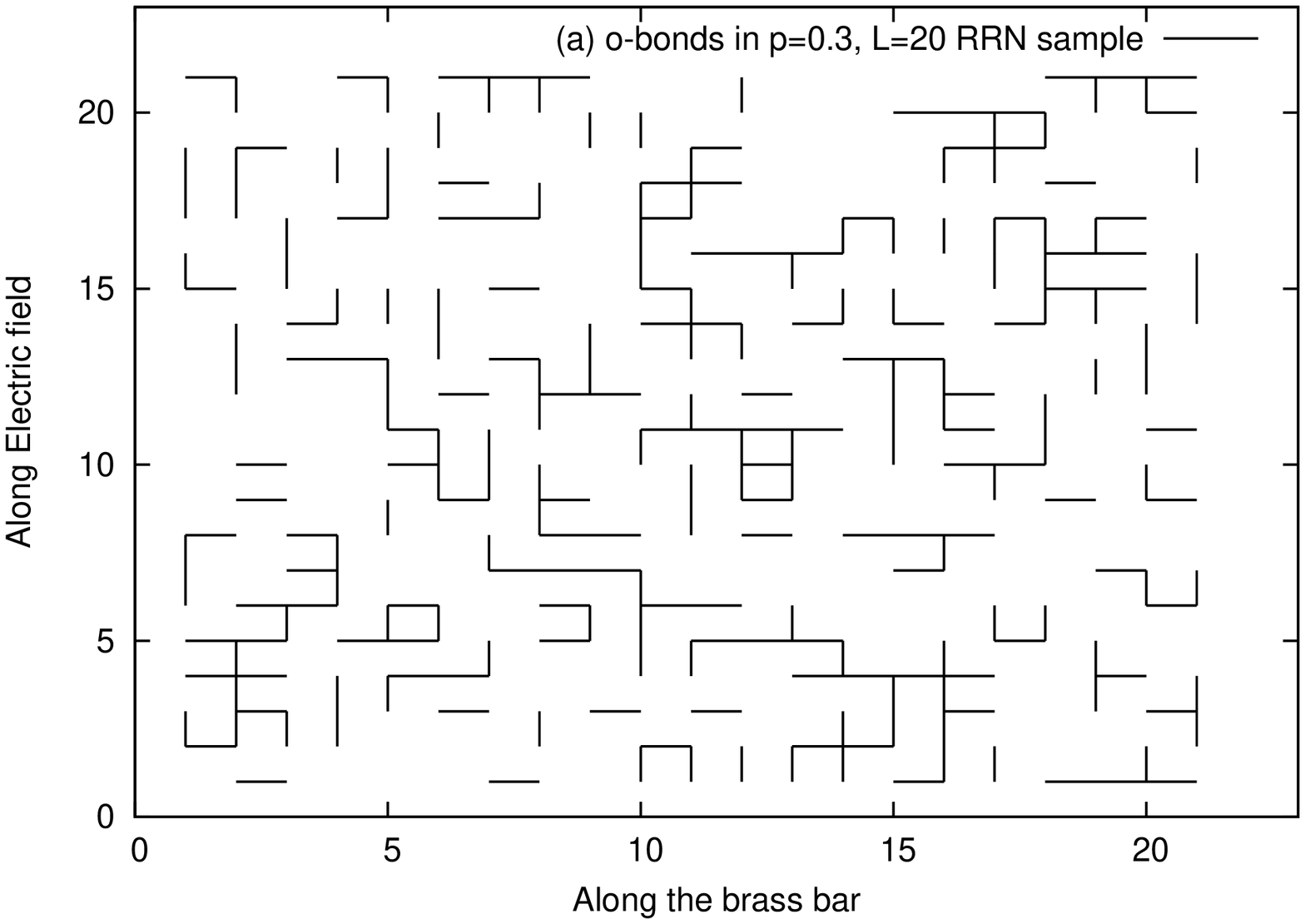}}}
\resizebox*{7cm}{7cm}{\rotatebox{270}{\includegraphics{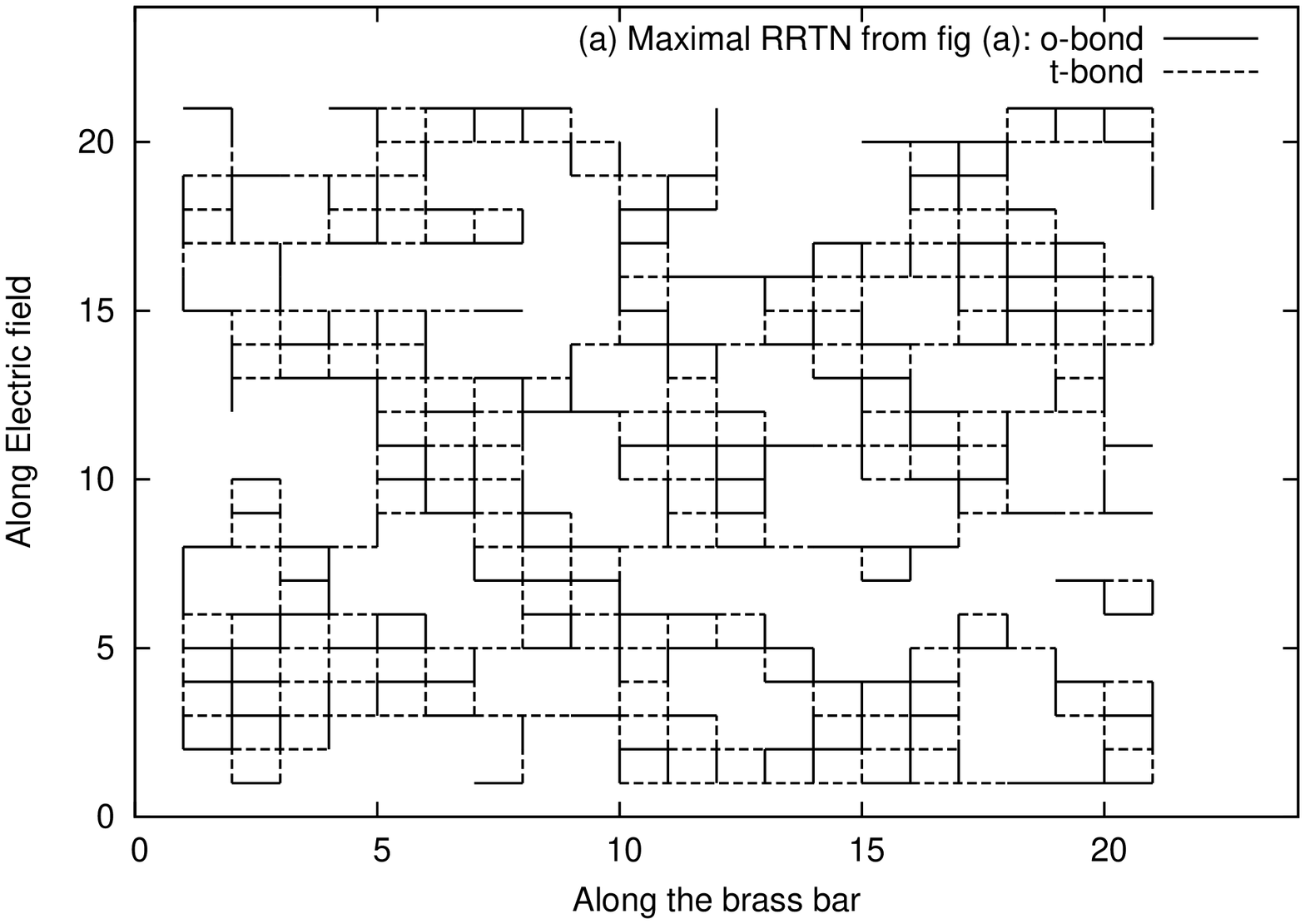}}}
\caption{(a) A non-percolating random o-bond configuration of a $L=20$, $p=0.30$ RRN square lattice.  The blank portion of the lattice has insulating bonds. (b) The {\it maximal} RRTN lattice is obtained from the given RRN [in fig. (a)] by considering the t-bonds (dashed lines) between any two nearest neighbour o-bonds. A t-bond becomes active when the potential difference across it is above a pre-assigned microscopic voltage threshold, \(v_g\).  This typical RRTN lattice in figure percolates by the help of active t-bonds.}
\label{lattice}
\end{figure}

\noindent  The electrical transport through metal-insulator binary composites is being investigated through various kinds of bond-percolation models for several decades. The current flow through these systems follow some tenuous system-spanning paths from one side of the material to the opposite side. For studying the linear transport phenomenon within the realm of statistical mechanics, one primarily considers a classical model like Random Resistor Network (RRN) \cite{stah} (See fig. [\ref{lattice}](a) for a representative RRN lattice), where the electrical conduction takes place {\it only} through some randomly arranged ohmic/linear resistors, known as ohmic bond (or o-bond). One may express the microscopic conductance of an ohmic bond as $g_o$. Afterwards some extra transport possibilities (e.g., like dielectric breakdown above a parametric threshold value of electric field) have been considered to investigate the nonlinear electrical response in a composite system. The Random Resistor cum Tunneling Network (RRTN) model, being a square-lattice semi-classical bond-percolation model, belongs to this class. In addition to the ohmic transport in RRN, this model considers an extra phenomenological reversible barrier-crossing process above a pre-assigned microscopic voltage threshold $v_g$ across \textit{some} insulating bonds. They are known as {\it tunneling} bonds (or t-bond). Here the phrase "microscopic" is used to describe a length scale much smaller than the sytem-size, but significantly larger than the atomic dimension. This RRTN model was primarily proposed \cite{rrtn}, to mimic the electrical transport through soft binary composite materials having ultra-low percolation threshold. With that these composites also lack any experimental response below a small macroscopic cut-in voltage, say $V_g$.

\noindent  Under the application of an external voltage, a composite material like every open system, needs some finite time to relax to a time-independent (or steady) current.  For studying the relaxation behaviour, one usually measures the bulk current, say, $I(t)$, as a function of time $t$, during its passage since the switching-on the external voltage to a steady state.  In major cases, this relaxation behaviour may be mathematically expressed in terms of a purely exponential function like, $I(t)= {\rm exp}(-t/\tau)$,  where $\tau$  is called the {\it relaxation time}. This is known as, Debye-class. Every linear system always relaxes this way. In systems, where nonlinear cooperative behaviour is dominant among the sub-systems, one generally finds a non-Debye type of relaxation behaviour, where $I(t)$ can not be mathematically expressed by a single exponential function. Power-law relaxation behaviour, experimentally observed in diverse types of composite materials (see references in \cite{sust}), belongs to an intriguing type within the non-Debye class.  

\noindent In literature, one may find different approaches to study the relaxation behaviour of binary-composite systems. We have followed a numerical approach based on \textit{iterative} Gauss-Seidel method to study the current relaxation for the RRTN, named  the method as {\it lattice Kirchhoff's dynamics} in refs. \cite{sust,epl}. Numerically to study any bulk electrical property of a discrete model, manifested from the cooperative behaviour of its microscopic sub-systems (e.g., like RRTN), one usually starts with an {\it initial} (i.e., guess) local voltage distribution within the lattice. For numerical study of the current dynamics, the choice of the initial voltage distribution is always arbitrary. In the further iterations, one needs to adopt an updating rule for the voltage at each node. Choice of the algorithm in this initial value problem depends on the relevant physical laws involved in the process. With that, it must assure the convergence of the iterative process. The bulk relaxation behaviour, observed experimentally for that network, can be theoretically obtained from the arithmetic average of (ergodic) dynamics, realised with different initial voltage configurations. For a set of microscopic voltages and conductances of the bonds, one may calculate the bulk current through the network for every iteration (time). As the microscopic voltages at each node change in time, so the bulk current through the network also varies in the consecutive iterations till the current becomes steady at asymptotically infinite time. 
%%%%%%%%%%%%%%%%%%%%%%%%%%%%%%%%%%%%%%%%%%%%%%%%%%%%%%%%%%%%%%%%%%%%%%%%%%%%%%%%%
%%%%%%%%%%%%%%%%%%%%%%%%%%%%%%%%%%%%%%%%%%%%%%%%%%%%%%%%%%%%%%%%%%%%%%%%%%%%%%%%%%%%
\noindent In our previous numerical investigations on RRTN bulk current relaxation since refs. \cite{sust,epl}, we usually consider a randomised graded initial voltage distribution (RGIV) at each node of the RRTN lattice. We observed a non-Debye type relaxation behaviour in early-time, followed by an asymptotic single exponential tail. More interestingly, for {\it some} of the RRTN dynamics, the non-exponential part can be expressed \cite{epl} as a couple of inverse power-laws (i.e., scale-free). Recently in ref. \cite{relax-physicaA}, we identified two time-scales (e.g., $\tau_t$ and $\tau_s$) for this dynamics. We studied the distribution of those time-scales within large number of RRTN samples for different system sizes. It was established that, out of them only one time-scale is linearly independent in the thermodynamic limit. 

\noindent In this present paper, we have investigated on the bulk current relaxation through the {\it first-passage} (FP) route in the RRTN model. In case of RRTN (whose RRN backbone is non-percolating), the first flush of macroscopic {\it finite} current (or equivalently, first-passage of the electrical response) is realised when a system-spanning conducting path is formed by the help of correlated activity of the t-bonds. Here the relaxation behaviour starts with an {\it additional} regime of almost zero (i.e., below detectable response) bulk current, before a sharp jump to a measurable value. This is followed by a non-exponential and then a Debye regime as well. The relaxation channel for which the explicit formation of percolating cluster is observed within the RRTN, is referred as first-passage route. Experimentally, the non-exponential relaxation behaviour have been reported in the responses of various binary-composite systems. We cite some recent works in multi-component electronic material \cite{rnp}, plasticized PMMA-LiClO$_4$ solid polymer electrolyte \cite{aswani}. This can also be observed in composites of cork granules and epoxy resin during mechanical stress relaxation \cite{reis}.

\noindent As per the identified physical origin of the time-scales like $\tau_t$ and $\tau_s$, they appear for every RRTN current relaxation. With that, one may identify an additional time-scale $\tau_B$, which corresponds to the first passage of bulk current in the FP route. So there are altogether three time-scales (e.g., $\tau_B$, $\tau_t$ and $\tau_s$), those evolve for the current dynamics in the FP route. Each of them generates due to the physically distinct process. However one may enquire that whether their values are correlated or not. We have worked on the distribution of the time-scales for a sufficiently large sample-sizes. We find that the ratios of the time-scales (e.g., $r_t \equiv \frac{\tau_t}{\tau_B}$ and $r_s \equiv \frac{\tau_s}{\tau_B}$), asymptotically converge to some constant values for the thermodynamic limit. From here we may claim that, there exists only one independent time-scale during any arbitrary kind of RRTN current dynamics. For systems whose bulk response is due to the non-gaussian diffusion of charge carriers, the first-passage time changes as a power-law with the system size, observed in biological network \cite{brown} and hydraulic flow in soil \cite{water}. For RRTN, the calculated power-law exponent describes a sub-diffusive behaviour.

\section{Model}

\noindent To capture the basic physics for some random mixtures of metal-insulator, structurally having very low percolation threshold of metallic part and mixtures which show their response above a small but finite threshold voltage, a square-lattice bond percolation model was proposed by Sen and Kar Gupta \cite{rrtn} decades ago. They named it as Random Resistor cum Tunneling Network (RRTN) model. It was necessary for such a new model to introduce a scope of electrical transport beyond the classical mechanism. For this purpose, in addition to the ohmic transport due to o-bonds, a threshold conduction process was introduced for some insulating bonds. This threshold conduction process was refered as "semi-classical tunneling" in the introductory reference \cite{rrtn}. In this model, this is implemented by assigning a microscopic voltage threshold of response ($v_g$) for each t-bond. Any insulating (dielectric) bond of the RRN, which is between any two o-bonds as nearest neighbour, is attributed as t-bond in the RRTN. A RRTN lattice, configured with all geometrically possible t-bonds, is referred as {\it maximal RRTN} (See fig. [\ref{lattice}](b)). The appearance of the t-bonds in this {\it perfectly correlated} (i.e., deterministic) fashion is the origin of a very low percolation threshold in the RRTN model. The percolation threshold of a $2D$ square lattice maximal RRTN (referred as, $p_{ct}$) is $0.181$  \cite{rrtn}. These t-bonds play as a conducting bridge between two neighbouring o-bonds, above a potential difference of $v_g$. The activation of a t-bond is physically possible due to the {\it reversible} dielectric breakdown at $v_g$. The microscopic conductance of an active t-bond is expressed as $g_t$. However below $v_g$, a t-bond remains inactive like an insulating material. This piece-wise linear response of a t-bond with different microscopic conductances on either sides of threshold voltage (i.e., $v_g$) is responsible for its nonlinear electrical behaviour. If the underlying RRN for a particular RRTN  sample is non-percolating, then that macroscopic RRTN will have a macroscopic threshold voltage $V_g$ (or equivalently a bulk breakdown voltage). For any finite external voltage ($V$) below $V_g$, no bulk steady current is (statiscally) expected. One finds that the bulk dc differential conductance $G (V) (\equiv dI/dV)$ vs. $V$, looks like a S-shaped (or sigmoidal) curve \cite{rrtn,dc}. It possesses two linear regimes at high and low $V$, known as {\it Upper Linear Regime} (ULR) and {\it Lower Linear Regime} (LLR). For the intermediate voltages, as more and more t-bonds start participating in the percolation cluster with the increase in $V$, so a nonlinear response is observed in the $G(V)$.  This region is termed as {\it Sigmoidal Regime}. It is gratifying for us that, in addition to the success of RRTN in explaining the nonlinear dc and ac response \cite{dc,ac}, the model can also reproduce several interesting aspects of dielectric breakdown \cite{break} and low-temperature variable range hopping \cite{vrh,ictp} as observed in case of composite materials.

%%%%%%%%%%%%%%%%%%%%%%%%%%%%%%%%%%%%%%%%%%%%%%%%%%%%%%%%%%%%%%%%%%%%%%%%%%%%%%%%%%%%%%%%%%%%%%%%%%%%%%%%%%%%%%%%%%%%%%%%%%%%%%%%%%%%%%%%%
\begin{figure}[tb]
\resizebox*{7cm}{7cm}{\rotatebox{270}{\includegraphics{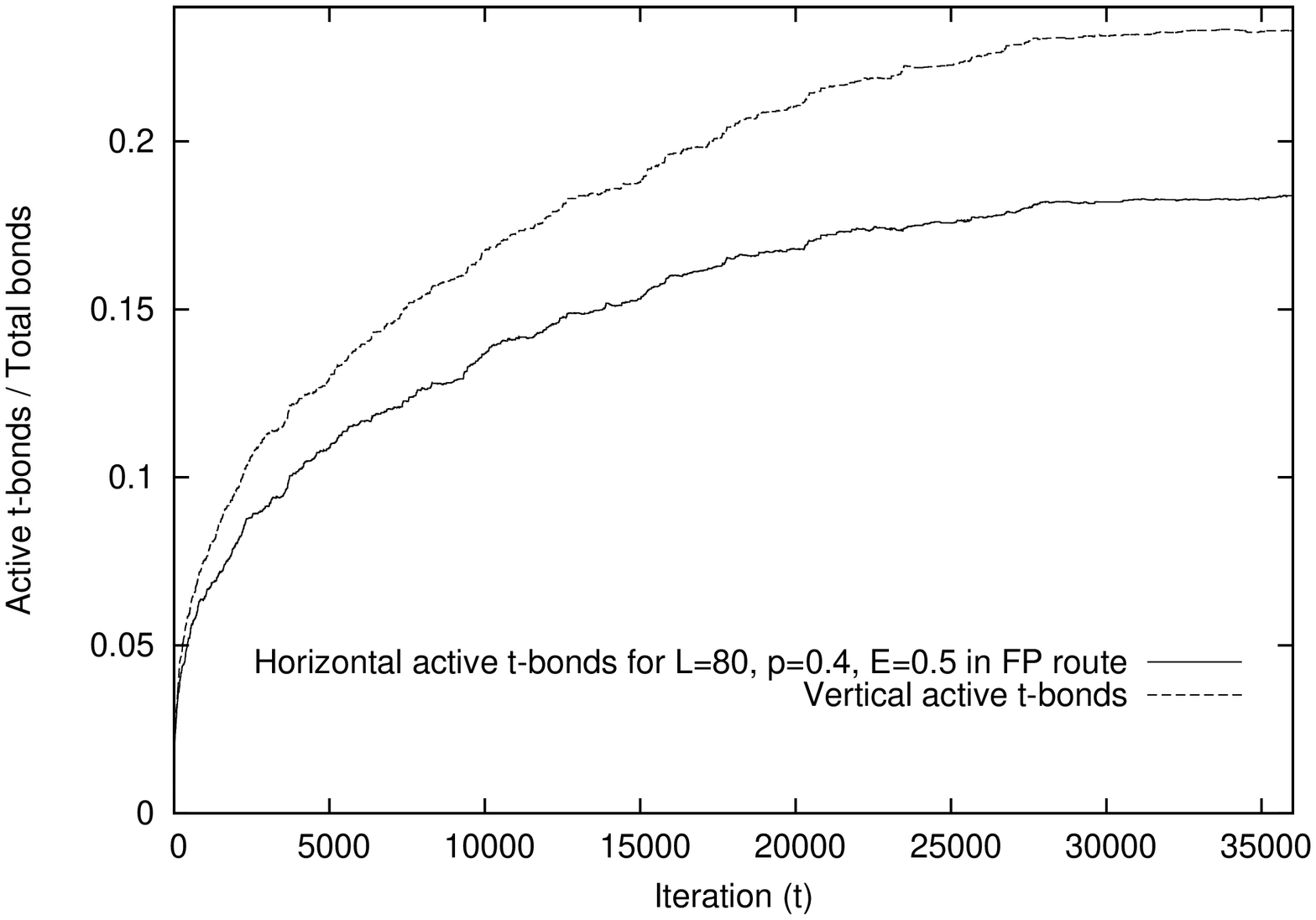}}} 
\resizebox*{7cm}{7cm}{\rotatebox{270}{\includegraphics{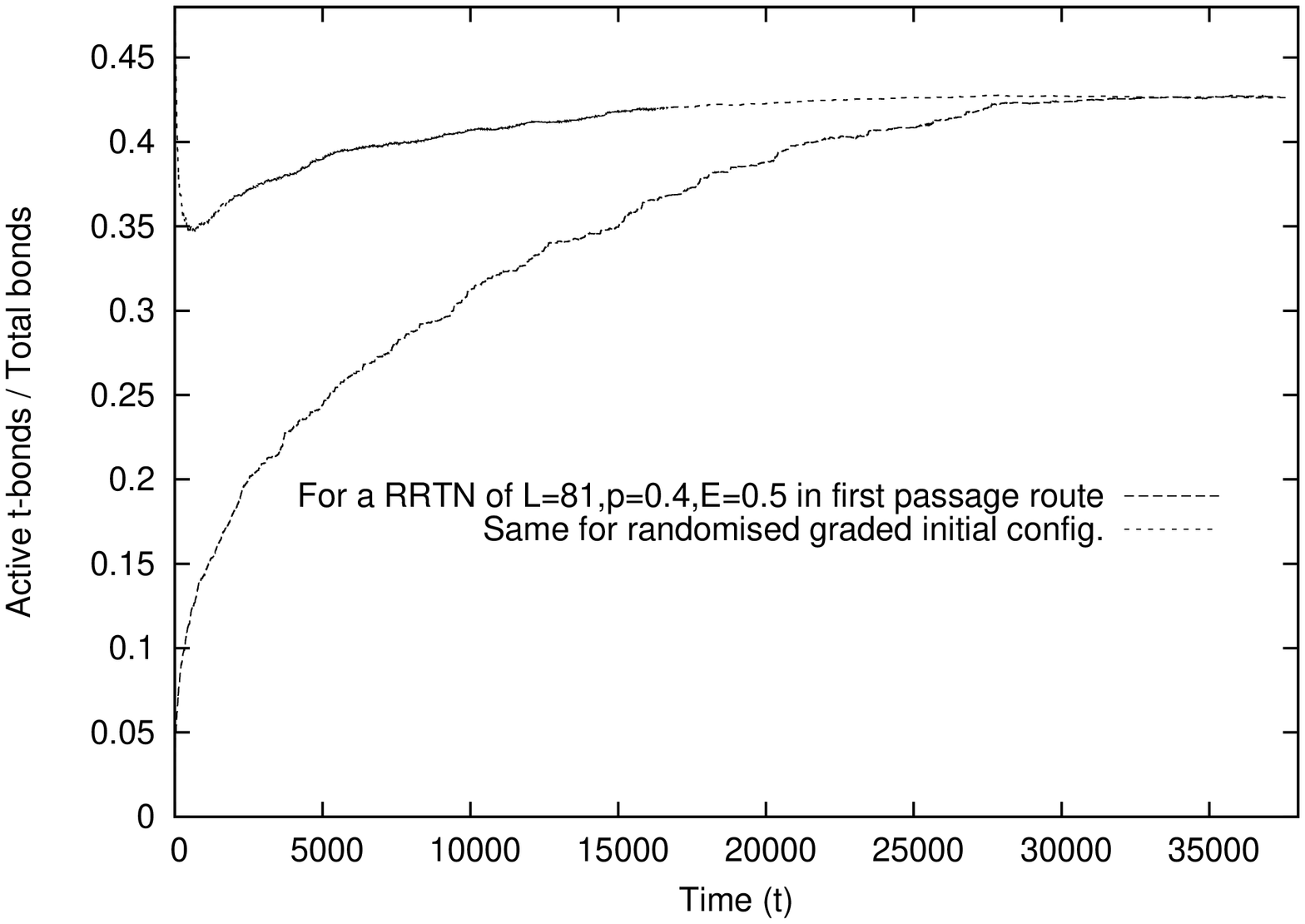}}}
\caption{(a) The number of vertical (along electric field) and horizontal (perpendicular to field) t-bonds for a RRTN sample of $L=80$, $p=0.4$ under the external field $E=0.5$, with iteraion (time). The number itself as well as the rate is faster for the vertical bonds than those for horizontal bonds. This figure classifies the growth of cluster as directed percolation.;(b) The variation of total no. of active t-bonds .vs. iteration time for the same RRTN lattice in case of current relaxation of first-passage (FP) route and randomised graded initial voltage (RGIV) configuration. In common for both cases, the number asymptotically becomes constant. It monotonically increases for the FP, whereas one observes an early-time dip for RGIV case.}
\label{t-bond-evolve}
\end{figure}
%%%%%%%%%%%%%%%%%%%%%%%%%%%%%%%%%%%%%%%%%%%%%%%%%%%%%%%%%%%%%%%%%%%%%%%%%%%%%%%%%%%%%%%%%%%%%%%%%%%%%%%%%%%%%%%%%%%%%%%%%%%%%%%%%%%%%%

\section{Kirchhoff's Dynamics}

In our numerical study, we apply a fixed voltage $V$ across RRTNs of different system sizes ($L$) and ohmic bond concentrations ($p$), as per the requirement. We study the time-evolution of the transient bulk current through a RRTN towards steady-state. The steady state of a network is realised, when the macroscopic current through it becomes constant in time. It means the aim of our dynamical study is to achieve the global current conservation for the system. In a system, where any macroscopic property evolves through the cooperative behaviour of the sub-systems, the steady-state of bulk current can only be maintained as an outcome of {\it simultaneous} local current conservation at each node of the lattice. We follow an approach based on the standard {\it Gauss-Seidel} (GS) method for our investigation. We aim to find the local voltage distribution at each node of the network, corresponding to local current continuity at each node. Because this will {\it only} assure the global steady state for a particular RRTN, as the charge density does not explicitly vary in time. The specific algorithm for our work, is termed as {\it lattice Kirchhoff's dynamics} since \cite{sust,epl}. The equations, we need to solve, are set accordingly, so that the {\it vectorial} sum of local currents, meeting at each node, vanishes. Writing the same in terms of local voltages and conductances, 

\begin{equation}
%\begin{displaymath}
\label{eq.1}
\sum\limits_{nn} g_{nn}\left(v_{nn} - v(j,k)\right) = 0, \forall (j,k). \hskip 2.0cm    
\end{equation}
%\end{displaymath}

\noindent
Here the sum has been taken over currents $i_{nn} = g_{nn}\left(v_{nn} - v(j,k)\right)$ through various types of nearest neighbour ($nn$) microscopic bonds around any arbitrary node $(j,k)$ of the lattice. $v(j,k)$ is the microscopic voltage at node $(j,k)$, 
$g_{nn}$ is the microscopic conductance of bond bridging a nearest neighbour site (with local voltage as $v_{nn}$) of node $(j,k)$. For the case of a square lattice, one considers the four $nn$'s around a node inside the bulk (three and two $nn$'s respectively at any boundary or a corner). A discrete, {\it scaled} time unit has been chosen as one completes scan through each site of the lattice. 

\noindent As we start with an arbitrary initial (or guess) local voltage distribution for RRTN nodes, the eq.~(\ref{eq.1}) will not hold with instantaneous values of local voltages (for most of the nodes) till the steady state. At time $t$ (within transient state), consider the non-zero value of total local current meeting at node $(j,k)$ is $\sum\limits_{nn} i_{nn}(t)$. It can be calculated from L.H.S. of eq.~(\ref{eq.1}), substituting local values at time $t$. Following GS method, an additive correction  term (ie., $\delta(j,k,t)$) to the instantaneous value of local voltage at node $(j,k)$ (i.e., $v(j,k,t)$) can be deduced, so that $v(j,k,t+1)$ {\it perfecty} satisfies eq.~(\ref{eq.1}) for the node $(j,k)$. During this propositon, it is pre-assumed that we have already substitued all the neighbouring local voltages of $v(j,k,t)$, i.e., \{$v_{nn}(t)\}$, with their corresponding steady-state values. Thus we write,

\begin{equation}
\sum\limits_{nn} g_{nn}(t) \left\{v_{nn}(t) - \left(v(j,k,t)+ \delta(j,k,t) \right)\right\} = 0. \hskip 1cm \forall (j,k) 
\label{eq.2}
\end{equation}

From eq.~(\ref{eq.2}), we find the correction term as, 

\begin{eqnarray}
\delta(j,k,t) &=& \frac{\sum\limits_{nn} g_{nn}(t)\left(v_{nn}(t) - v(j,k,t)\right)} { \sum\limits_{nn} g_{nn}(t)}, \nonumber \\
&=& \frac{ \sum\limits_{nn} i_{nn}(t)}{ \sum\limits_{nn} g_{nn}(t)}.
\end{eqnarray}

Thus the updating rule for lattice Kirchhoff's dynamics will be,
 
\begin{equation}
{v(j,k,t+1)} = {v(j,k,t)} + \frac{\sum\limits_{nn} i_{nn}(t)} {\sum\limits_{nn} g_{nn}(t)}   \hskip 1cm \forall (j,k).
\label{update}
\end{equation}

\noindent One may note, as the microscopic conductance of a t-bond depends on the instantaneous local voltage-difference across it, so in principle the conductance of any neighbouring bond i.e., $g_{nn}(t)$ becomes time-dependent. On the other hand, the $\{v_{nn}(t)\}$ values during transient state, practically differ from their steady-values. So even after the initiaive in eq.~(\ref{update}), one needs further refinement to attain steady-state local voltage distribution. Thus, we numerically solve the set of coupled equations on the RRTN lattice in an iterative way following an algorithm, designed for successive refinement. On a phenomenological ground, the move towards a macroscopic steady-state implies that the difference of currents through the first and the last layers tends to a pre-assigned smallness as a function of time. We remember that \cite{relax-physicaA}, for any kind of arbitrariness both in initial conditions of $\{v(j,k)\}$ as well as in the choice of RRTN lattice, the algorithm is strongly convergent . 

\noindent In this paper, we investigate on the current relaxation during the first-passage route for the RRTN. To accomplish this task numerically, we have connected one brass-bar (say, first horizontal layer of the network) with uniform voltage bias $V$ and the other brass-bar (say, last horizontal layer) to ground (i.e., zero voltage). Now to set up the microscopic {\it initial} voltage configuration inside the RRTN sample appropriately, voltages at all the internal nodes of RRTN (i.e., $v(j,k,0) \forall j~,~k=2,L $) are assigned to zero value as well. We update the microscopic internal voltages for next iterations till steady state following eq. (\ref{update}). For our numerical calculation, we use the parameter values as $g_o=1.0$ for each o-bond and $g_t = 10^{-2}$ with $v_g=0.5$ for each t-bond (all in some arbitrary units). 
%%%
%%%%%%%%%%%%%%%%%%%%%%%%%%%%%%%%%%%%%%%%%%%%%%%%%%%%%%%%%%%%%%%%%%%%%%%%%%%%%%%%%%%%%%%%%%%%%%%%%%%%%%%%%%%%%%%%%%%%%%%%%%%%%%%%%%%%%%
\begin{figure}[htb]
\centering
\resizebox*{6cm}{6cm}{\rotatebox{270}{\includegraphics{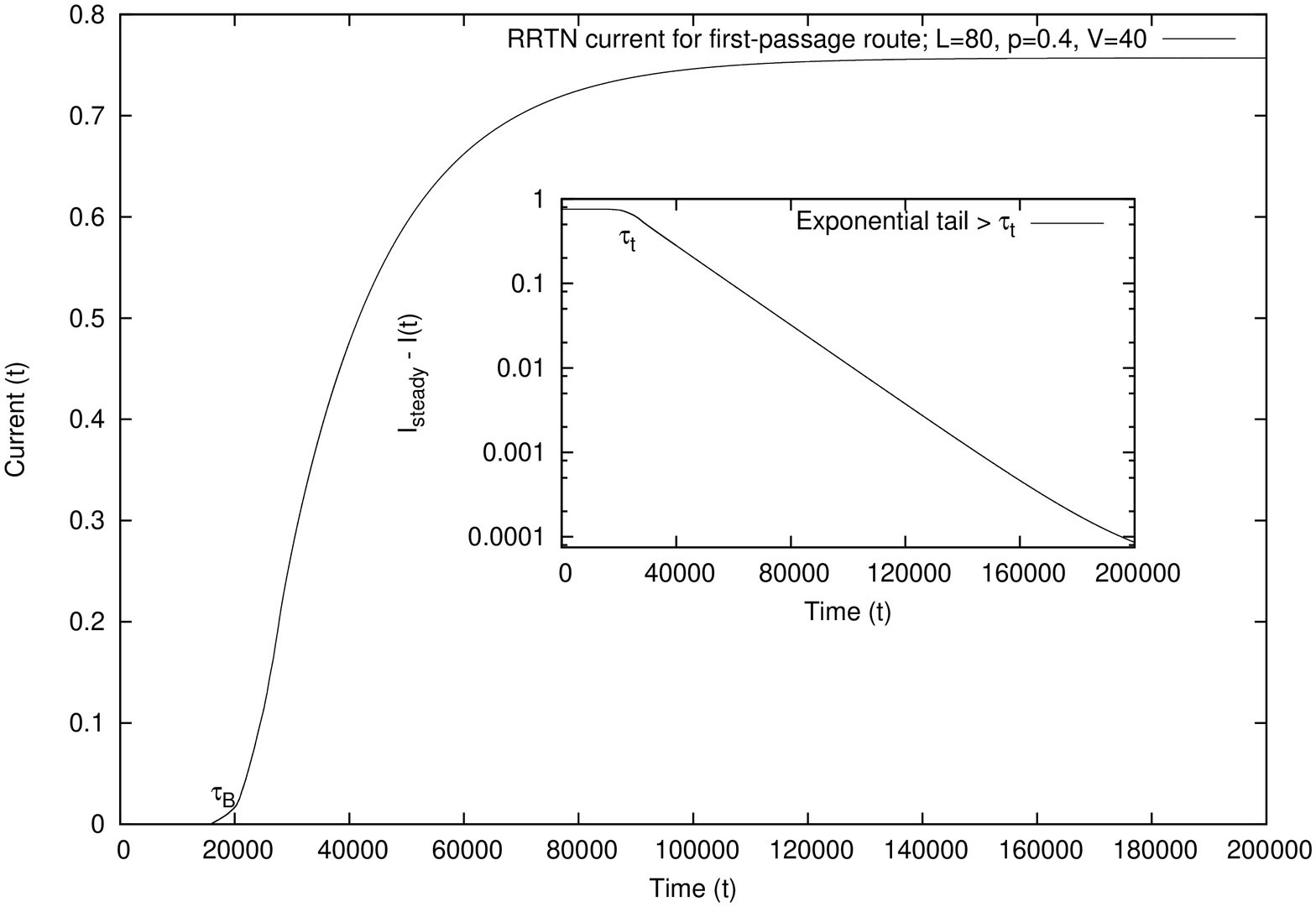}}}
\resizebox*{6cm}{6cm}{\rotatebox{270}{\includegraphics{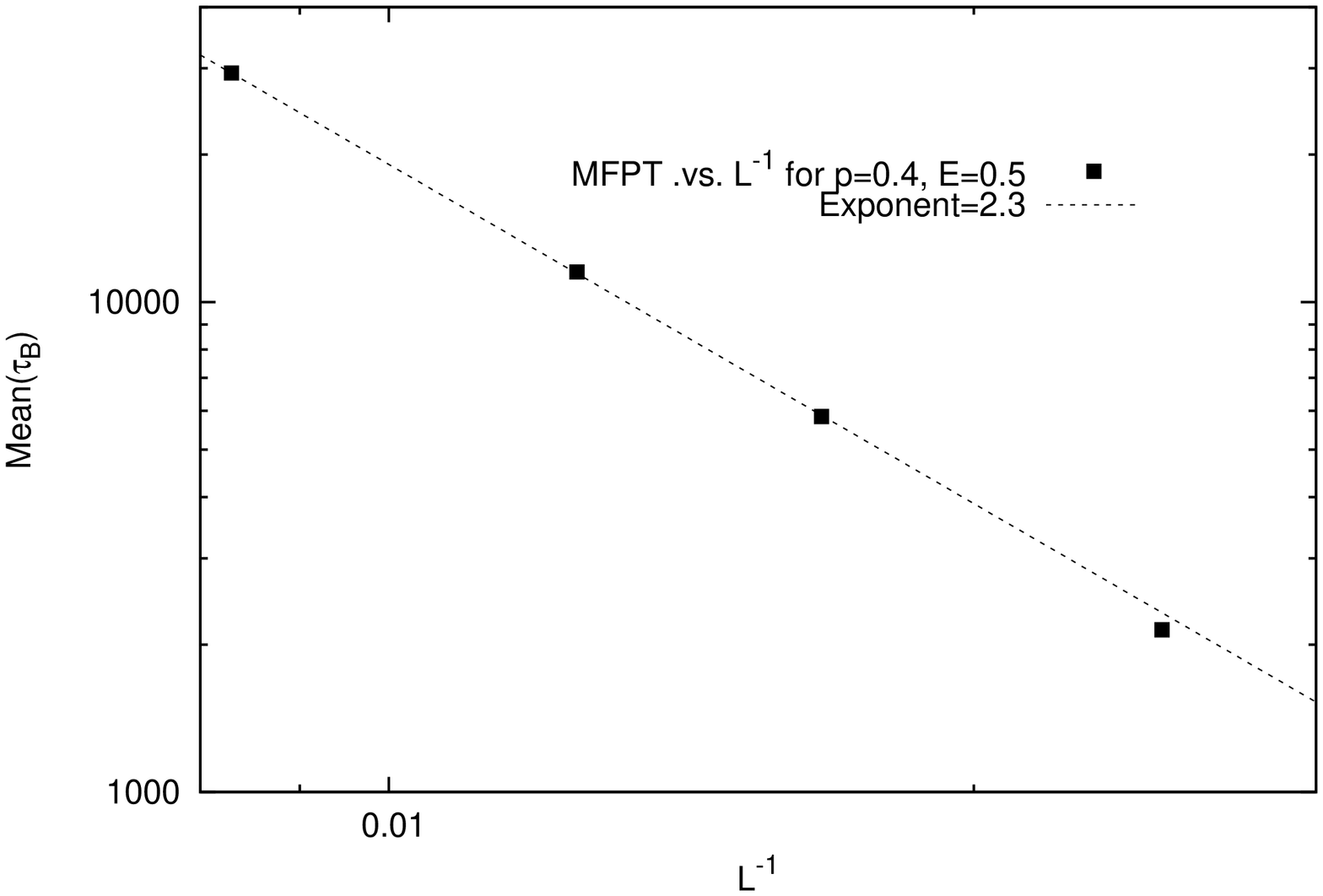}}}
\caption{(a) A current-relaxation dynamics of the same RRTN for the first-passage route. The bulk current $I(t)$ does not show any measurable value till a breakdown time $\tau_B$, then there is a non-exponential growth in the current upto the time $\sim \tau_t$. Finally, it approaches to its steady-state (upto $\tau_s$) through a purely Debye behaviour (shown in inset where a semi-log scale is chosen to plot [$I(steady)-I(t)$]). The time-scales which describe the extent of the regimes have been labelled.; (b)The variation of mean first-passage time ($\langle \tau_B \rangle$) through RRTN with corresponding system sizes $L$. We have fitted with a power-law function as, $\langle \tau_B \rangle \sim L^\alpha$. Our data averaged over $5000$ RRTN samples of $p=0.4, E=0.5$ fits with $\alpha = 2.3$.}
\label{fraction}
\end{figure}
%%%%%%%%%%%%%%%%%%%%%%%%%%%%%%%%%%%%%%%%%%%%%%%%%%%%%%%%%%%%%%%%%%%%%%%%%%%%%%%%%%%%%%%%%%%%%%%%%%%%%%%%%%%%%%%%%%%%%%%%%%%%%%%%%%%%%%
%%%%%%%%%%%%%%%%%%%%%%%%%%%%%%%%%%%%%%%%%%%%%%%%%%%%%%%%%%%%%%%%%%%%%%%%%%%%%%%%%%%%%%%%%%%%%%%%%%%%%%%%%%%%%%%%%%%%%%%%%%%%%%%%%%%%%%%%%
\begin{figure}[htb]
\resizebox*{7cm}{7cm}{\rotatebox{270}{\includegraphics{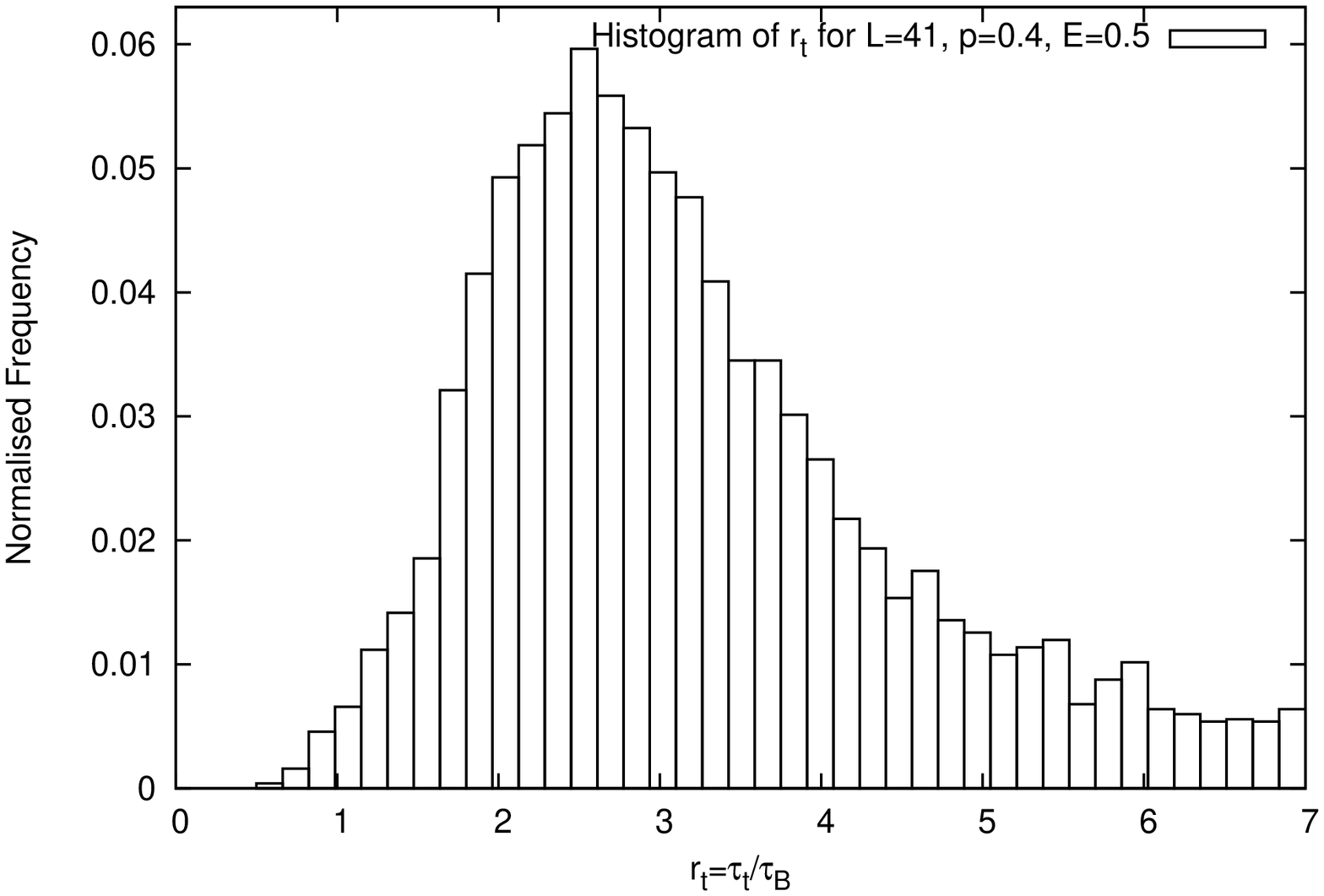}}}
\resizebox*{7cm}{7cm}{\rotatebox{270}{\includegraphics{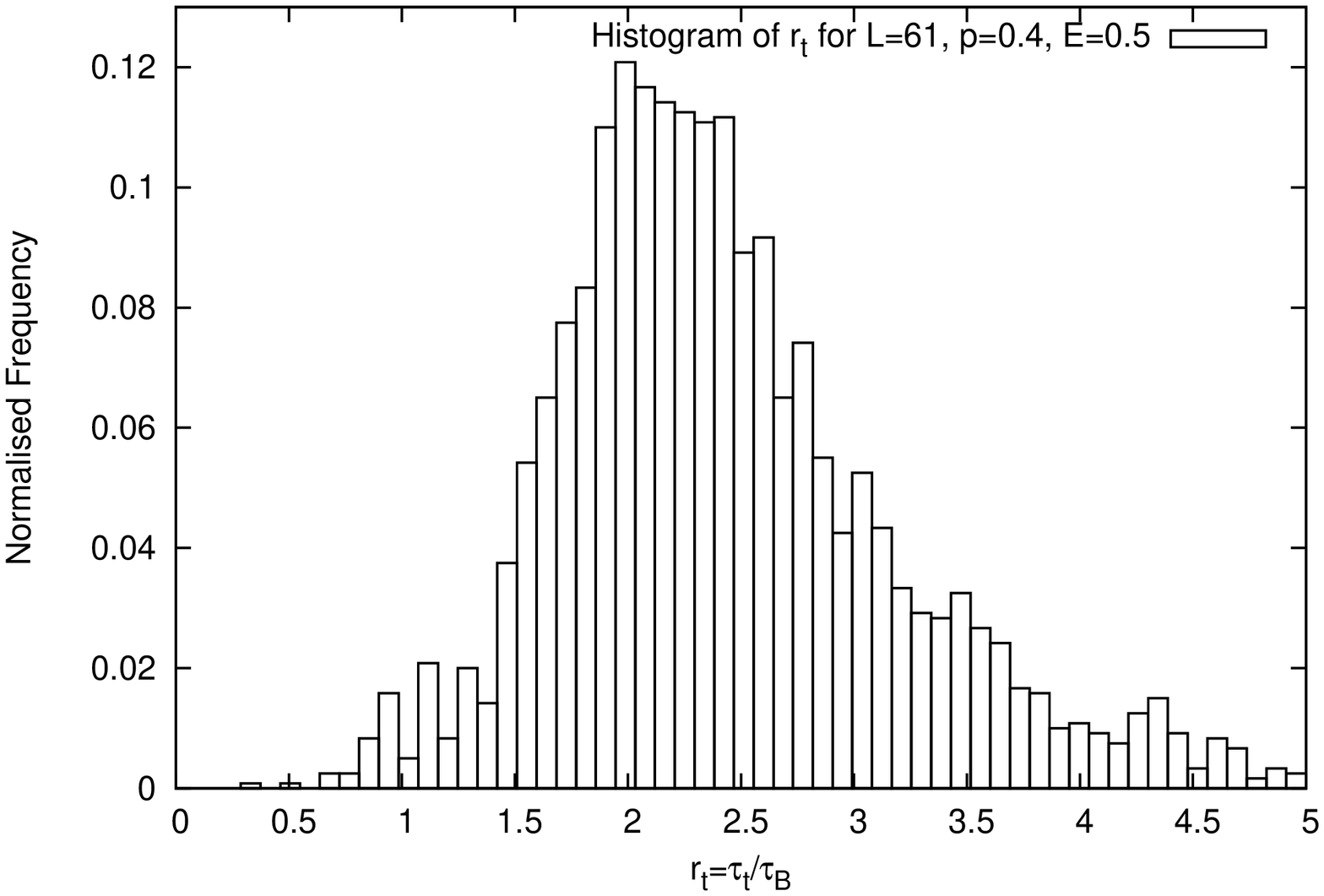}}} 
\caption{(a) The histogram for the ratio of the time-scales (i.e., $\tau_t/\tau_B$) using $N=5000$ random RRTN bond-configurations with system-size $L=40$, ohmic concentration $p=0.4$ for average electric field $E=0.5$. By {\it Normalised frequency}, we mean the actual frequency divided by total number of configurations used i.e., $N$. The behaviour possesses a prominent peak $r_t(40)$ around $2.53$; (b) A similar histogram for RRTN samples with system size $L=60$, showing its mode $r_t(60)$ around $1.98$.}
\label{hisrat1}
\end{figure}
%%%%%%%%%%%%%%%%%%%%%%%%%%%%%%%%%%%%%%%%%%%%%%%%%%%%%%%%%%%%%%%%%%%%%%%%%%%%%%%%%%%%%%%%%%%%%%%%%%%%%%%%%%%%%%%%%%%%%%%%%%%%%%%%%%%%%%%%%
%%%%%%%%%%%%%%%%%%%%%%%%%%%%%%%%%%%%%%%%%%%%%%%%%%%%%%%%%%%%%%%%%%%%%%%%%%%%%%%%%%%%%%%%%%%%%%%%%%%%%%%%%%%%%%%%%%%%%%%%%%%%%%%%%
\begin{figure}[htb]
\resizebox*{7cm}{7cm}{\rotatebox{270}{\includegraphics{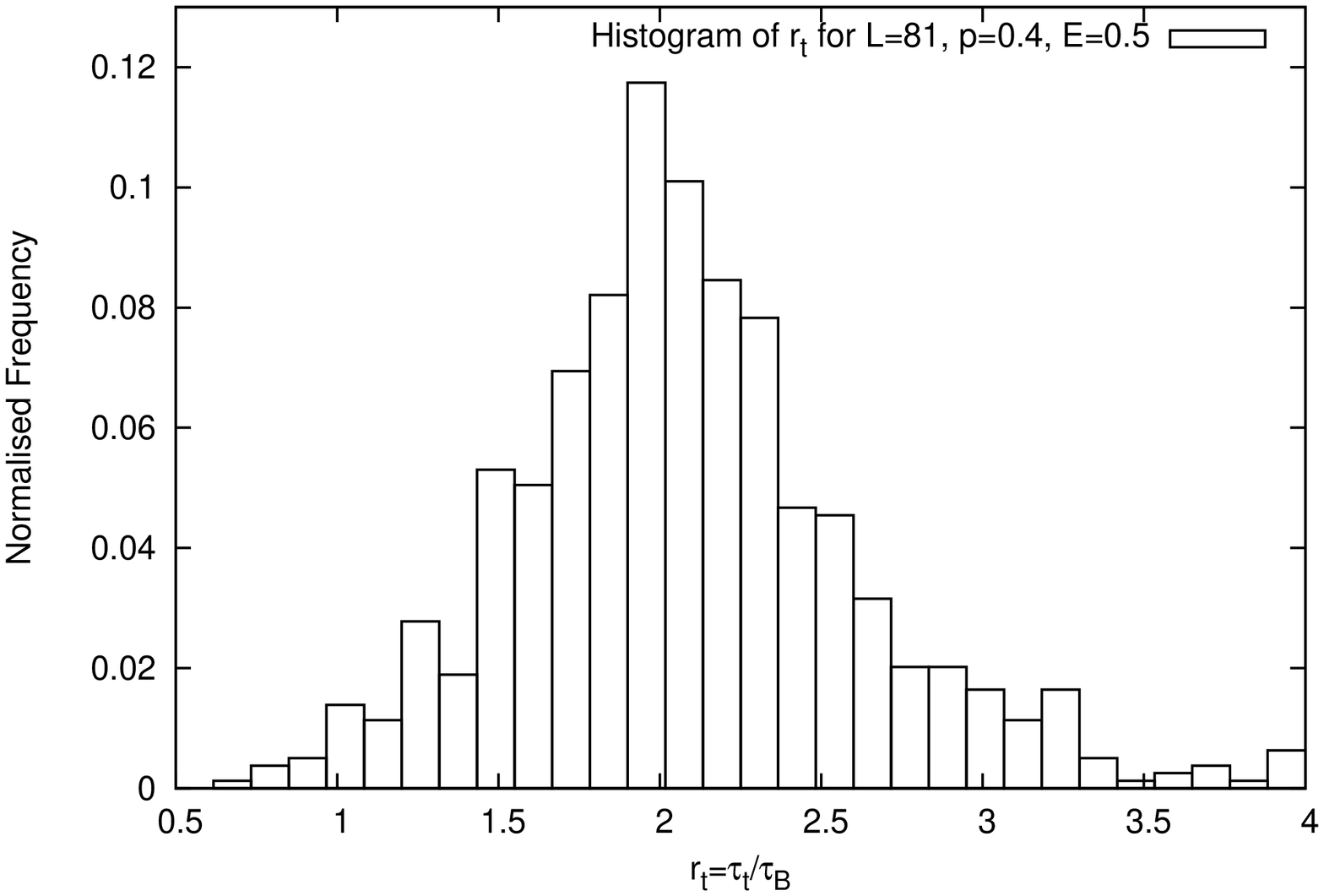}}}
\resizebox*{7cm}{7cm}{\rotatebox{270}{\includegraphics{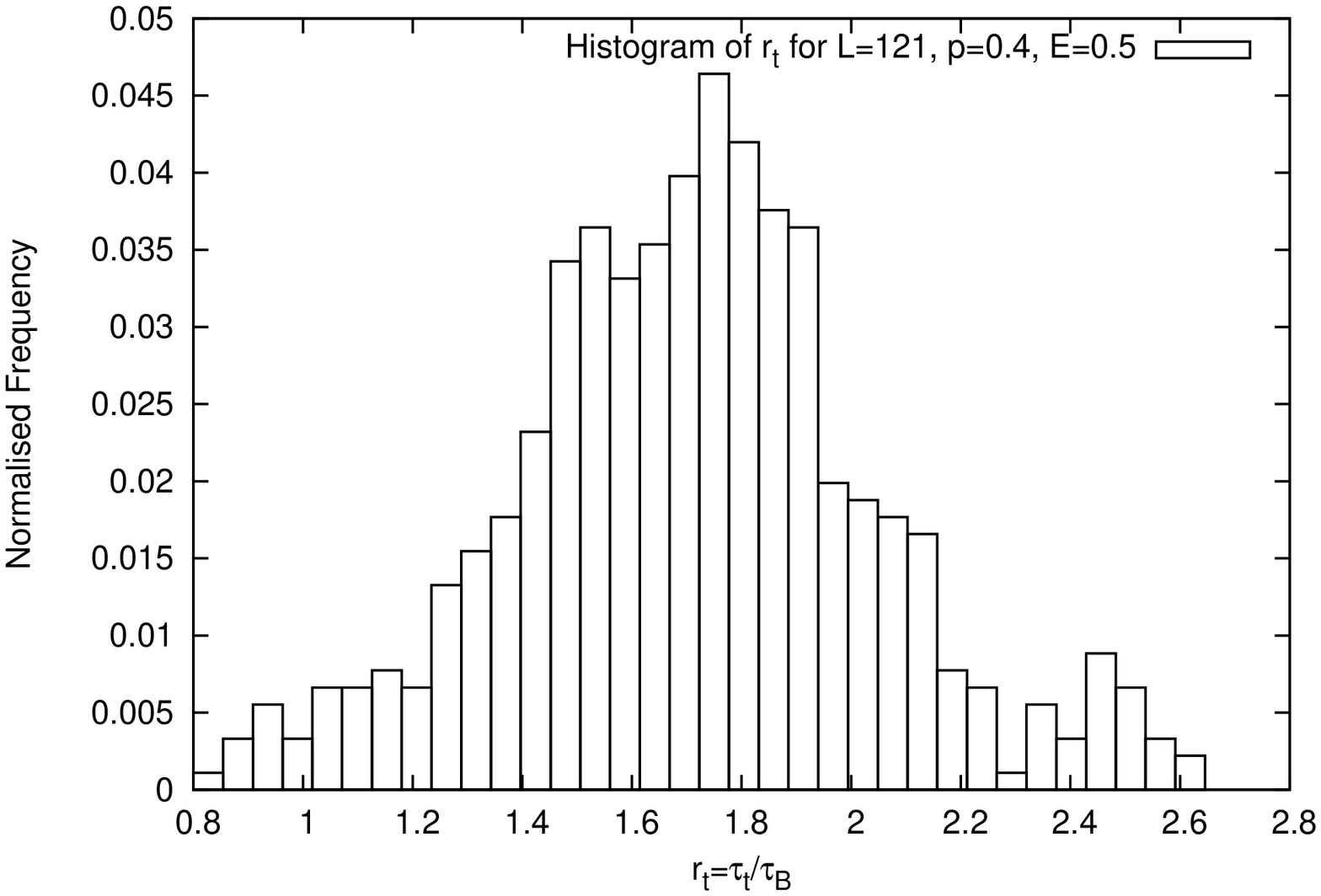}}}
\caption{(a) A histogram for RRTN samples with system size $L=80$, showing mode $r_t(80)$ around $1.95$. There is a sharp peak, almost symmetric around its mode.; (b) A similar histogram for RRTN samples with system size $L=120$, showing a mode $r_t(120)$ around $1.75$. }
\label{hisrat2}
\end{figure}
%===========================================================
%%%%%%%%%%%%%%%%%%%%%%%%%%%%%%%%%%%%%%%%%%%%%%%%%%%%%%%%%%%%%%%%%%%%%%%%%%%%%%%%%%%%%%%%%%%%%%%%%%%%%%%%%%%%%%%%%%%%%%%%%%%%%%%%%%%%%%%%%

\section{Results and Discussions}
 
\noindent In this paper, we mainly focus on the bulk current relaxation during the first passage (FP) route in the RRTN network. First, we mention about our motivation of studying the current dynamics in this route. Then we qualitatively describe the corresponding temporal regimes of the RRTN current relaxation and discuss on their physical origins. One may here identify three time-scales (expressed as $\tau_B$, $\tau_t$ and $\tau_s$), which qualitatively indicate the extent of regimes during the dynamics. Finally, we investigate on the distribution-statistics of time-scales using several ($\sim 5000$) RRTN samples of different $(L,p,V)$. We observe a perfect mutual correlation among them in the thermodynamic limit. This enables us to comment that only one time-scale, out of three, is independent. Analysing the mean first-passage time, we find the movement of charge-carrier within the RRTN as similar to sub-diffusive motion.

\subsection{Motivation of studying RRTN current relaxation in the first-passage route}

\noindent The bulk current ($I$) through an open electrical system under fixed external voltage ($V$) relaxes to a steady value (say, $I_{steady}$) in course of time ($t$) after the bias is switched-on. At every time, the value of macroscopic relaxing current (i.e., $I(t)$) through the system purely depends on the contribution of the microscopic currents from different sub-systems. Those local currents add up vectorially to generate the global current. In the non-steady state of relaxation, each local current value keeps on varying with time due to the change in the local voltages. For a square-lattice bond percolation model like RRTN, local currents are the currents through different bonds and the local voltages are the microscopic voltages at different nodes (i.e., $(j,k)$) of the network.  The bond current can be obtained by multiplying the bond-conductance with the voltage difference across the bond. Thus essentially for every time, we need to input the complete set of voltages at each node for calculating the currents through RRTN bonds. At steady state, the local currents and voltages for a fixed network (like maximal RRTN) can only be determined analytically using Kirchhoff's law. But in case of a RRTN, where the network itself (or equivalently the participation of t-bonds) depends on the local voltages, no exact analytical calculation of the local voltage is possible for any arbitrary time. So the numerical approach becomes inevitable for finding the local currents within RRTN. 

\noindent To initiate the RRTN current dynamics, one needs to start with some initial (i.e., guess) local voltage distribution. Once a voltage configuration is available, one may adopt a logical rule to update the set for use in the next iteration and follow the same procedure repetitively for the further times. Thus, setting up a voltage configuration initially plays a crucial role in the investigation of the bulk current relaxation for any network as the calculation of bond-currents entirely depends on this choice. Unfortunately there is no standard procedure to determine an initial voltage configuration for a disordered network, except it must not violate the principle of "action-at-a-distance". Thus in general, one  chooses an initial set of voltages {\it arbitrarily} so that the dynamics reaches to a {\it convergent} steady state. In our earlier papers \cite{sust, epl}, we started the current dynamics with a randomised graded initial microscopic voltage (RGIV) distribution. The graded voltage distribution expresses the exact steady local voltages when the network is perfectly ordered (means $p=1.0$) with all bonds as ohmic resistors with equal conductance value.  We randomise over this graded voltage distribution to initiate the RRTN current dynamics. For that we use pseudo-random numbers which are uniformly distributed within $[0,1]$.

\noindent While starting the dynamics with RGIV configuration, we {\it pre-assigned} a finite voltage at each internal node of the RRTN sample and oblige the network to relax from there. Though due to our choice any layer closer to the positive brass-bar, is {\it probabilistically} being assigned to a higher voltage (i.e., action-at-a-distance is obeyed), but with that all internal nodes also attain some finite voltages in no time. This means the effect of external voltage percolates instantaneously, just after it is connected with RRTN. We observe a finite global current since the first iteration time. During the relaxation behaviour with RGIV, we missed to observe the sequential building of system-spanning percolation cluster with the help of t-bonds. Thus the phenomenon like first-passage of bulk current was also absent during the relaxation. 

\noindent The internal dynamics during the formation of percolating cluster (of conducting bonds) may provide useful information of transport-mechanism through a network. In this present paper, we intend to study the RRTN bulk current dynamics via the first-passage route. After being connected to an external source, the network distributes the local voltages by its own, gradually during the relaxation. In this route, as the system is directly in touch with the external voltage, so only the terminal layers (i.e., nodes with $k=1,L ~\forall j$) will instantaneously attain the source voltages. On the other hand, the voltages at each internal layers of the RRTN will be updated sequentially obeying eq. [\ref{update}]. Till then, they do not get any initial voltages to start with. One may note here that, in eq. [\ref{update}], the updation of any $v(j,k)$ {\it explicitly} involves the local conductances of its neighbouring RRTN bonds (i.e., those meeting at node $(j,k)$ ), as well as the effect of local conductances enters {\it implicitly} during the calculations of local currents through those RRTN bonds. Thus in this route of relaxation, one observes a network-specific spreading of the effect of external voltage across the RRTN and first flush of bulk current when the network percolates for the first-time. 

\subsection{Regimes of RRTN current relaxation in the first-passage route}

\noindent The effect of the external voltage continuously spreads internally during the relaxation process in first-passage (FP) route. It is appropriate to mention, the RRN backbone of any RRTN lattice does not change during the entire relaxation process. The instantaneous network changes at every instant only due to the freshly active t-bonds. Thus the formation of RRTN percolation cluster may be equivalently studied through following the time-series of the number of active t-bonds (i.e., for those $|v| \geq v_g$). In fig. [\ref{t-bond-evolve}](a), we show that the number of both vertical (along electric field direction) and horizontal (along perpendicular to electric field) t-bonds for a representative RRTN sample of $L=80$, $p=0.4$ under the external field $E=0.5$, simultaneously grow in course of iteraion (time). Over the square-lattice, the network can grow only along these two directions. So  increase in number of both kinds of t-bonds (with time) definitely establishes the formation of RRTN percolative cluster.  Moreover the growth-rate as well as the number of vertical bonds at any arbitrary time are higher than those of horizontal bonds. This behaviour describes a directed (i.e., towards the applied field) spreading of percolation cluster inside the RRTN network. In fig. [\ref{t-bond-evolve}](b), we demonstrate the time-variation of total number of active t-bonds during current-relaxation for FP route as well as for RGIV distribution. As common behaviour, the number saturates almost to a same constant value asymptotically after a finite time. This finite time is identified as one of the relevant time-scales. Then beyond this, the bulk current relaxation is purely Debye-type (with a single time-constant) till the current through the sample becomes steady. The time corresponding to the steady-state is another time-scale during relaxaion. In ref. \cite{relax-physicaA}, we named these two time-scales as $\tau_t$ and $\tau_s$ respectively. During the entire current-dynamics in FP route, no. of active t-bonds monotonically increases from zero since the first iteration. We may explain that, as all the internal nodes of RRTN are initially at zero voltage, so no t-bond is active. As time progresses, the effect of the external voltage gradually percolates inside, causing more and more further activation of t-bonds within the network. Due to this, one observes the formation of local-clusters, which further link up together to develop a system-spanning percolating-cluster, which is responsible for bulk constructive (dielectric) breakdown inside the sample. As the RRTN percolates, we notice the first-passage of bulk current. One may identify a process-specific time-scale $\tau_B$, corresponding to the onset of finite measurable current. The activation of t-bonds continues till $\tau_t$. To summarise, we observe three temporal regimes during the current relaxation in the FP route, viz, i) initially a pre-first-passage regime with no measurable current, characterised by the time-scale $\tau_B$, ii) then a finite response till around the time-scale $\tau_t$, where the system-current relaxes in a non-exponential manner, iii) finally an asymptotic exponential tail towards the steady-state, expressed by the time-scale $\tau_s$. In fig. [\ref{fraction}](a), we present the corresponding relaxation behaviour of macroscopic current during FP route. In the inset of the same figure, we plot the $I(steady)-I(t)$ in a semi-log scale, to present the exponential variation during the growth of current. Interestingly we note, the time scales during FP route are directly associated with three major bulk mechanisms during relaxation, viz., constructive fracture, active participation of nonlinear agents (or t-bonds) and establishment of the global continuity in carrier-current.

\noindent In contrary, during the relaxation with RGIV distribution, the number of active t-bonds sharply decreases from an initial (i.e., at $t=0$) finite number to a minimum for the first few hundreds of iteration-times. Then there is a rise in its value till the number becomes asymptotically constant. Actually after the RGIV is initially {\it imposed} on the internal nodes for a percolating RRTN, the current relaxation starts in presence of pre-assigned sufficiently exess number of active t-bonds within a fixed RRN lattice. Due to this, one observes a finite initial macroscopic current through RRTN, which decreases for the entire current-dynamics towards steady-state. However in this stuation, the local continuity condition becomes {\it highly} imbalanced. Then to adjust the local current continuity, a fast de-activation of t-bonds (as a reactive mechanism) takes place during early-times. Due to strong relaxing-inertia, the deactivation process of t-bonds continues in such a way that, inside RRTN the number of active t-bonds undershoots to a smaller value. It stops to a minimum. Further due to the self-organisation, the number monotonically increases till $\tau_t$. Thus the distribution process of the active t-bonds in this route is not a purely directed diffusion upto $\tau_t$ as observed in case of FP route. 

%%%%%%%%%%%%%%%%%%%%%%%%%%%%%%%%%%%%%%%%%%%%%%%%%%%%%%%%%%%%%%%%%%%%%%%%%%%%%%%%%%%%%%%%%%%%%%%%%%%%%%%%%%%%%%%%%%%%%%%%%%%%%%%%%%%%%%%%%
\begin{figure}
\centering
\resizebox*{6cm}{6cm}{\rotatebox{270}{\includegraphics{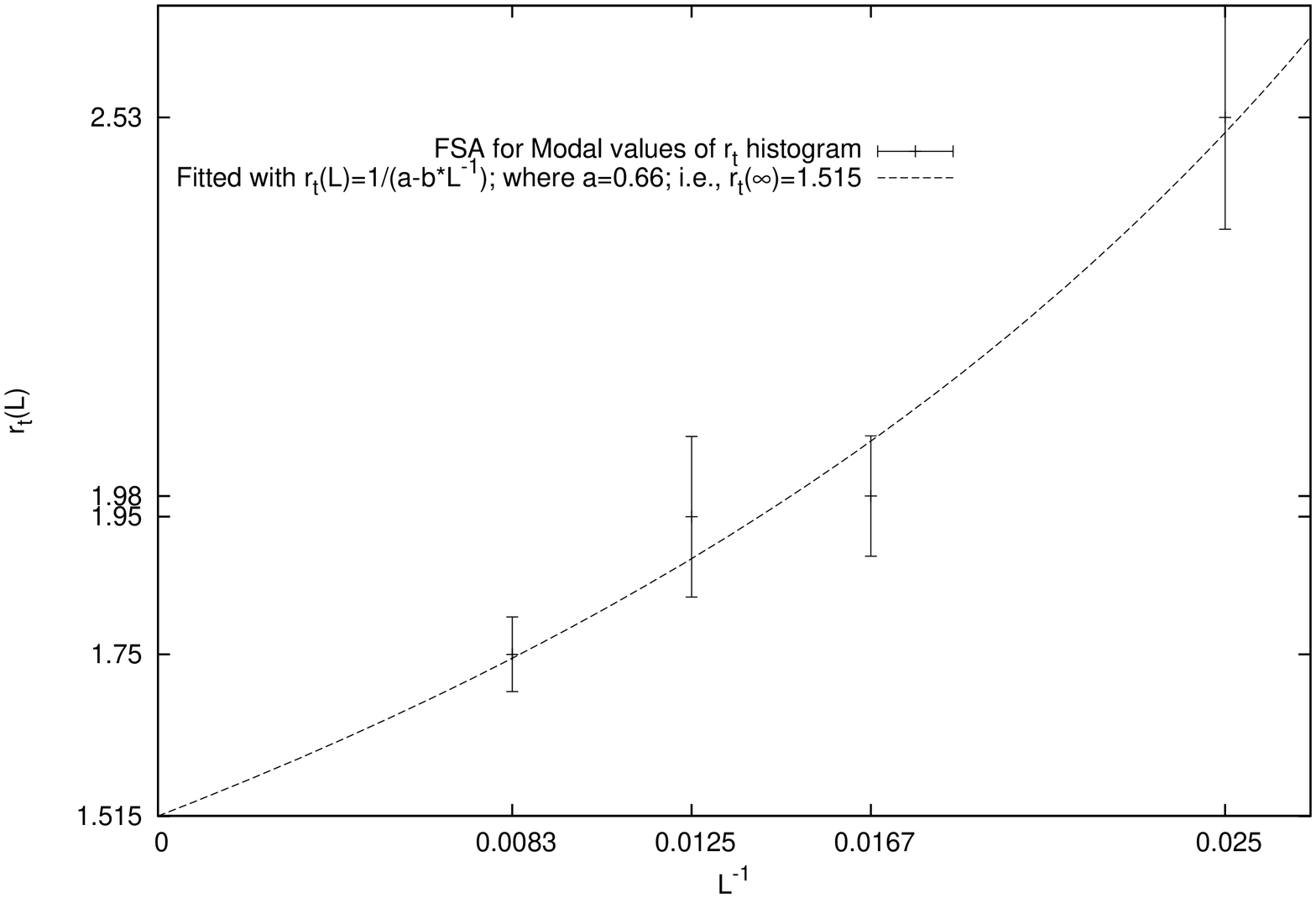}}}
\resizebox*{6cm}{6cm}{\rotatebox{270}{\includegraphics{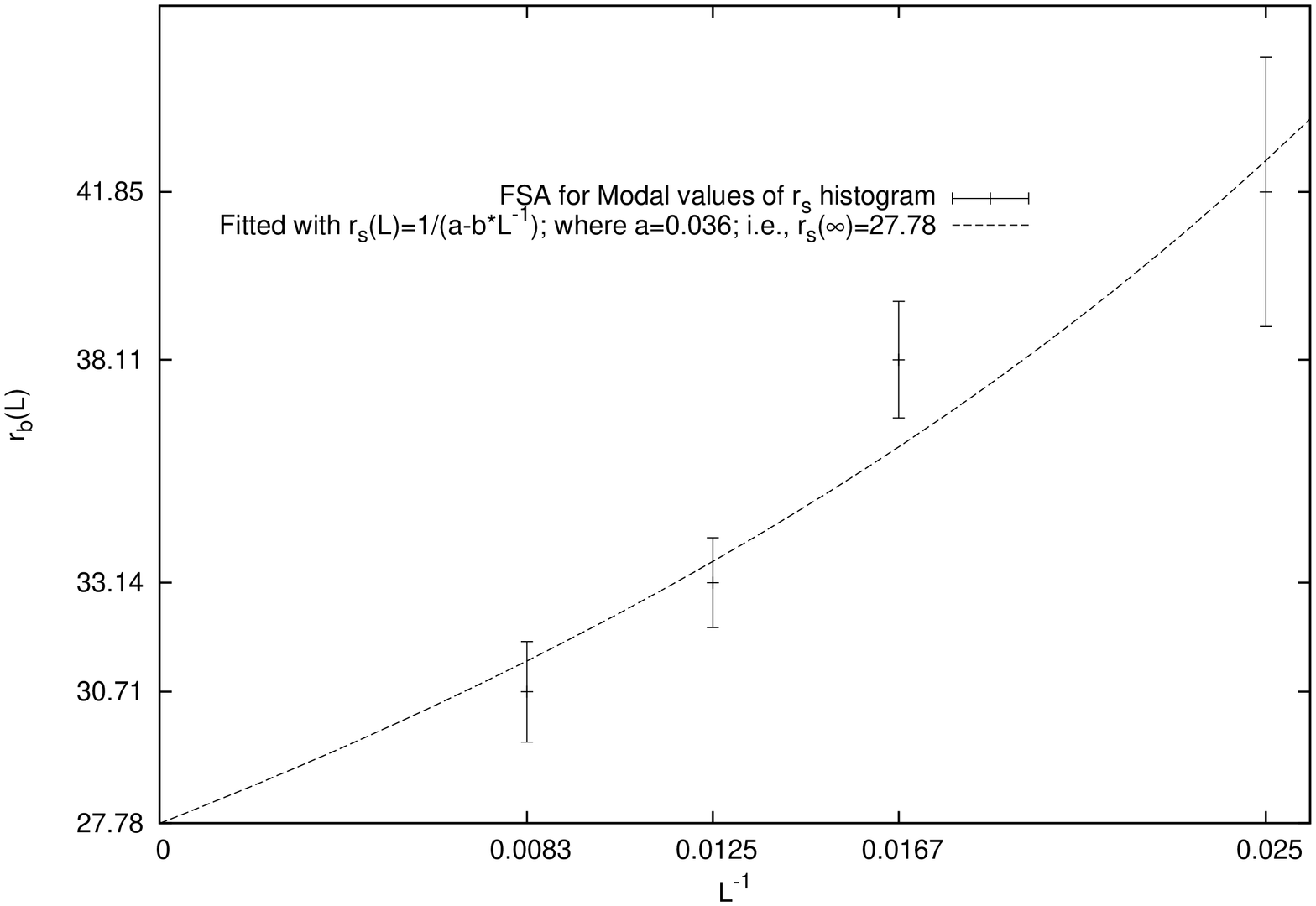}}}
\caption{(a) A graph showing the finite size analysis (FSA) on the mode values of the previous histograms (i.e., figs. [\ref{hisrat1},\ref{hisrat2}]) using an extrapolation formula, $r_t(L)=\frac{1}{a-b~L^{-1}}$. On fitting, we find, $r_t(\infty)=a^{-1} \sim 1.515$.; (b) A graph showing the finite size analysis (FSA) on the mode values of the $r_s$ histograms (not included in this paper) using an extrapolation formula, $r_s(L)=\frac{1}{a-b~L^{-1}}$. On fitting, we find, $r_s(\infty)=a^{-1} \sim 27.78$.}
\label{fsa}
\end{figure}
%%%%%%%%%%%%%%%%%%%%%%%%%%%%%%%%%%%%%%%%%%%%%%%%%%%%%%%%%%%%%%%%%%%%%%%%%%%%%%%%%%%%%%%%%%%%%%%%%%%%%%%%%%%%%%%%%%%%%%%%%%%%%%%%%%%%%%%%%

\subsection{Correlation among the time-scales, $\tau_B$, $\tau_t$ and $\tau_s$ }

\noindent  During the entire RRTN current-relaxation process through the FP route, one may identify three relevant time-scales viz., $\tau_B$, $\tau_t$ and $\tau_s$. These time-scales quantitatively describe the extent of broad regimes of the current dynamics, as mentioned in detail in the previous subsection. Two of them (e.g., $\tau_t$ and $\tau_s$) were identified in our recent paper \cite{relax-physicaA}, where we also detailed on their quantitative estimation. These two time-scales have a universal origin, in the sense that they originate for any arbitrary route of current relaxation in RRTN. However, the time $\tau_B$, that scales the first measurable bulk response, {\it only} appears in FP route. The symbol $\tau_B$ was coined in an elementary study \cite{aksubh} from our group. To estimate $\tau_B$ in our present numerical work, we define it as the iteration-time, when the bulk current is just above a pre-assigned finite value (Say, $10^{-16}$ (in some arbitrary unit)). We have measured each of the time-scales for a set of $5000$ random RRTN bond-configurations, each with ohmic concentration $p=0.4$ under electric field $E=0.5$. The task is repeated for the system sizes $L=20 - 120$. From there, we have calculated the mean first-passage time (i.e., $\langle \tau_B \rangle$) by appropriate sample-averaging on $\tau_B$. We plot the variation of $\langle \tau_B \rangle$ with system size $L$ and fit with a power-law function as, $\langle \tau_B \rangle \sim L^\alpha$. In fig. [\ref{fraction}](b), we find that the exponent is $\alpha = 2.3$. This power-law behaviour may generally appear due to a fractional diffusion process \cite{fdiff1,fdiff2}, as manifested in diverse types of disordered networks, even beyond physical systems \cite{brown,water}. Interestingly, Havlin et. al. \cite{hav-avra} for the anomalous-diffusion through a 2-d fractal medium (viz., Sierpinski Gasket) reported on r.m.s diffusion length after $N$-steps as, $(R_N)_{r.m.s} \sim N^{1/d_w}$ with $d_w=2.32 \pm 0.01$, which is pretty close with our fitted value of the exponent $\alpha$. 

\noindent The first-passage time ($\tau_B$) for bulk current through a RRTN, measures the time needed for a charge-carrier to complete its journey across that particular sample for the first-time. As the $\tau_B$ values calculated for different RRTN lattices with same $(p,L)$ shows a distribution, so we work with ensemble averaged quantity like $\langle \tau_B \rangle$. The mean square displacement for a brownian / diffusive particle increases linearly with time elapsed. In comparison, for RRTN $\alpha > 2$ can be understood due to sub-diffusive behaviour (i.e., slower than diffusive) for charge-carrier. Physically we may expect sub-diffusive  carrier motion within a RRTN (with non-percolating RRN as backbone) due to inevitable active participation of t-bonds in forming the percolating cluster. These bonds phenomenologically mimic hopping between trap / bound states during electrical transport \cite{vrh,ictp}. A t-bond participates only in threshold conduction, otherwise assists in local charge accumulation by insulating the local current to pass through it. A RRN percolates only by the help of o-bonds, where each of the bonds is purely diffusive in nature.    

\noindent On change of context, we note that the physical origins of the time-scales may undoubtedly differ, but it is necessary to investigate whether there exists any correlation among them or not. To verify that, the ratios of the time-scales like  $r_t \equiv \frac{\tau_t}{\tau_B}$ and $r_s \equiv \frac{\tau_s}{\tau_B}$ have been calculated from there and we plot the corresponding probability density functions in form of histograms. In figs. [\ref{hisrat1},\ref{hisrat2}], we show those histograms for $L=40, 60, 80, 120$.  Each histogram has a prominent peak with permissible width. We study the variation of the modes of each histogram (denoted by $r_t(L)$) (see captions of figs. [\ref{hisrat1},\ref{hisrat2}] for values) with corresponding system sizes $L$. We observe an asymptotic trend of saturation in the values of $r_t(L)$ at large $L$ limit. We have used a standard finite size analysis (FSA) to extrapolate the mode value in the thermodynamic limit (i.e., for $L \to \infty$). To fit the data, we use a typical formula like, $r_t(L)=\frac{1}{a-b~L^{-1}}$ and obtain $r_t(\infty)=1.515$ (shown in fig. [\ref{fsa}](a)) for the data in use. 

\noindent We repeat the same study for the ratio $r_s$. We observe that the mode values of the $r_s$ histograms (i.e., $r_s(L)$) also saturates for $ L \to \infty$. With use of an identical fitting function as before, we find $r_s(\infty)=27.78$. This study is presented in fig. [\ref{fsa}](b). Our study confirms that though one needs three time-scales to quantify the regimes of RRTN current relaxation in the first-passage route, but their values (in the thermodynamic limit) are purely correlated among each other, because they have constant ratios for large system sizes. This means, only one time-scale is independent, out of them.

\section{Conclusion}

\noindent We have studied the relaxation of the macroscopic current via the first-passage route for the RRTN model under a uniform external voltage. This study enables us to study the current dynamics during the formation of the RRTN percolating cluster for first-flush of response. We identify three broad temporal regimes of the relaxation. After an initial behaviour of almost zero bulk current, a first-passage of response is observed due to a system-spanning dielectric breakdown. We observed during the growth of percolating cluster, the carrier transport is like a sub-diffusive mechanism, as realised from the variation of $\langle \tau_B \rangle$ .vs. $L$. The measurable current relaxes non-exponentially, then crosses over to an asymptotic purely Debye tail with a single time constant till the response becomes steady. The non-Debye behaviour in the dynamics is generated solely due to the {\it transient} appearance of the active t-bonds within RRN skeleton. Eventually the Debye relaxation behaviour emerges when the total number of active t-bonds does not further change in time. The extensions of the regimes can be quantitatively expressed through three distinct time-scales (i.e., $\tau_B$, $\tau_t$ and $\tau_s$). Though the origin of each of them are qualitatively different, but we establish that quantitatively these time-scales are purely correlated on each other, as their ratios sufficiently converge to some constant values in the thermodynamic limit. Thus, the knowledge of each one of them will enable us to predict the values of the other two. So there exists only one {\it independent} time-scale during a RRTN current dynamics. One may note here, the current relaxation in first-passage route takes place between the instant of the system-spanning dielectric-breakdown and the time when the bulk-response becomes steady. The times which scale them are $\tau_B$ and $\tau_s$. On the other hand, the dynamics of the t-bonds in the RRTN (expressed through the variation of instantaneous number of active t-bonds) runs for a duration, which scales as $\tau_t$. As the bulk current flow in RRTN directly depends on the activity of t-bonds, so these two dynamics must mutually affect each other. But our most intriguing observation for this communicaton is, these time-scales are even {\it purely} correlated.

\acknowledgments
The author acknowledges the full support by UGC Minor Research Grant File No. {\it PSW-162/14-15 (ERO) 
dt. 3.2.15} in completion and presentation of the work. He also thanks the sincere cooperations from the postgraduate 
department of Physics, Barasat Govt. College, Kolkata, for providing the best academic logistics, during his past service tenure there.

\end{document}